\documentclass[12pt]{article}

\usepackage{amsmath,amsfonts,amssymb}

\begin{document}

\begin{titlepage}

\begin{center}

\begin{center}
{\Large{ \bf Uniform Gauge for D1-brane in General  Background}}
\end{center}

\vskip 1cm

{\large Josef Kluso\v{n}$^{}$\footnote{E-mail: {\tt
klu@physics.muni.cz}} }

\vskip 0.8cm

{\it Department of
Theoretical Physics and Astrophysics\\
Faculty of Science, Masaryk University\\
Kotl\'{a}\v{r}sk\'{a} 2, 611 37, Brno\\
Czech Republic\\
[10mm]}

\vskip 0.8cm

\end{center}

\begin{abstract}
We construct uniform gauge D1-brane action in general background. We
also discuss how this action transforms under double Wick rotation
and determine transformation properties of background fields.

\end{abstract}

\bigskip

\end{titlepage}

\newpage

\newcommand{\mT}{\mathcal{T}}
\def\mK{\mathcal{K}}
\def\tr{\mathrm{Tr}}
\def\bB{\mathbf{B}}
\def\tJ{\tilde{J}}
\def\str{\mathrm{Str}}
\newcommand{\tL}{\tilde{L}}
\def\Pf{\mathrm{Pf}}
\def\I{\mathbf{i}}
\def\IT{\I_{\Phi,\Phi',T}}
\def \cit{\IT^{\dag}}
\def \cdt{\overline{\tilde{D}T}}
\def \dt{\tilde{D}T}
\def\bra #1{\left<#1\right|}
\def\ket #1{\left|#1\right>}
\def\vac #1{\left<\left<#1\right>\right>}
\def\pb  #1{\left\{#1\right\}}
\def \uw #1{(w^{#1})}
\def \dw #1{(w_{#1})}
\newcommand{\bK}{\mathbf{K}}
\newcommand{\thw}{\tilde{\hat{w}}}
\newcommand{\bA}{{\bf A}}
\newcommand{\bd}{{\bf d}}
\def\bX{\mathbf{X}}
\newcommand{\bD}{{\bf D}}
\def\tsigma{\tilde{\sigma}}
\def\ttau{\tilde{\tau}}
\def\bV{\mathbf{V}}
\newcommand{\bF}{{\bf F}}
\newcommand{\bN}{{\bf N}}
\newcommand{\hp}{\hat{p}}
\newcommand{\hq}{\hat{q}}
\newcommand{\hF}{\hat{F}}
\newcommand{\hG}{\hat{G}}
\newcommand{\hH}{\hat{H}}
\newcommand{\hU}{\hat{U}}
\newcommand{\mH}{\mathcal{H}}
\newcommand{\mG}{\mathcal{G}}
\newcommand{\mA}{\mathcal{A}}
\newcommand{\mD}{\mathcal{D}}
\newcommand{\tpr}{t^{\prime}}
\newcommand{\bzg}{\overline{\zg}}
\newcommand{\of}{\overline{f}}
\newcommand{\ow}{\overline{w}}
\newcommand{\htheta}{\hat{\theta}}
\newcommand{\opartial}{\overline{\partial}}
\newcommand{\hd}{\hat{d}}
\newcommand{\halpha}{\hat{\alpha}}
\newcommand{\hbeta}{\hat{\beta}}
\newcommand{\hdelta}{\hat{\delta}}
\newcommand{\hgamma}{\hat{\gamma}}
\newcommand{\hlambda}{\hat{\lambda}}
\newcommand{\hw}{\hat{w}}
\newcommand{\hN}{\hat{N}}
\newcommand{\onabla}{\overline{\nabla}}
\newcommand{\hmu}{\hat{\mu}}
\newcommand{\hnu}{\hat{\nu}}
\newcommand{\ha}{\hat{a}}
\newcommand{\hb}{\hat{b}}
\newcommand{\hc}{\hat{c}}
\newcommand{\com}[1]{\left[#1\right]}
\newcommand{\oz}{\overline{z}}
\newcommand{\oJ}{\overline{J}}
\newcommand{\mL}{\mathcal{L}}
\newcommand{\oh}{\overline{h}}
\newcommand{\oT}{\overline{T}}
\newcommand{\oepsilon}{\overline{\epsilon}}
\newcommand{\tP}{\tilde{P}}
\newcommand{\hP}{\hat{P}}
\newcommand{\talpha}{\tilde{\alpha}}
\newcommand{\uc}{\underline{c}}
\newcommand{\ud}{\underline{d}}
\newcommand{\ue}{\underline{e}}
\newcommand{\uf}{\underline{f}}
\newcommand{\hpi}{\hat{\pi}}
\newcommand{\oZ}{\overline{Z}}
\newcommand{\tg}{\tilde{g}}
\newcommand{\tK}{\tilde{K}}
\newcommand{\tj}{\tilde{j}}
\newcommand{\tG}{\tilde{G}}
\newcommand{\hg}{\hat{g}}
\newcommand{\htG}{\hat{\tilde{G}}}
\newcommand{\hX}{\hat{X}}
\newcommand{\hY}{\hat{Y}}
\newcommand{\bDi}{\left(\bD^{-1}\right)}
\newcommand{\hthteta}{\hat{\theta}}
\newcommand{\hB}{\hat{B}}
\newcommand{\tlambda}{\tilde{\lambda}}
\newcommand{\thlambda}{\tilde{\hat{\lambda}}}
\newcommand{\tw}{\tilde{w}}
\newcommand{\hJ}{\hat{J}}
\newcommand{\tPsi}{\tilde{\Psi}}
\newcommand{\cP}{{\cal P}}
\newcommand{\tphi}{\tilde{\phi}}
\newcommand{\tOmega}{\tilde{\Omega}}
\newcommand{\homega}{\hat{\omega}}
\newcommand{\hupsilon}{\hat{\upsilon}}
\newcommand{\hUpsilon}{\hat{\Upsilon}}
\newcommand{\hOmega}{\hat{\Omega}}
\newcommand{\bJ}{\mathbf{J}}
\newcommand{\olambda}{\overline{\lambda}}
\newcommand{\uhlambda}{\underline{\hlambda}}
\newcommand{\uhw}{\underline{\hw}}
\newcommand{\tC}{\tilde{C}}
\def \lhw #1{(\hw^{#1})}
\def \dhw #1{(\hw_{#1})}
\newcommand{\bG}{\mathbf{G}}
\newcommand{\bhG}{\hat{\bG}}
\newcommand{\bH}{\mathbf{H}}
\newcommand{\bE}{\mathbf{E}}
\newcommand{\mJ}{\mathcal{J}}
\newcommand{\mY}{\mathcal{Y}}
\newcommand{\mZ}{\mathcal{Z}}
\newcommand{\hj}{\hat{j}}
\newcommand{\bAi}{\left(\bA^{-1}\right)}
\newcommand{\hpartial}{\hat{\partial}}
\newcommand{\hD}{\hat{D}}
\newcommand{\mC}{\mathcal{C}}
\newcommand{\omC}{\overline{\mC}}
\newcommand{\mP}{\mathcal{P}}
\newcommand{\omP}{\overline{\mP}}
\newcommand{\tLambda}{\tilde{\Lambda}}
\newcommand{\tPhi}{\tilde{\Phi}}
\newcommand{\tD}{\tilde{D}}
\newcommand{\tgamma}{\tilde{\gamma}}
\section{Introduction and Summary }
It is well known that the central role in our understanding of the
AdS/CFT correspondence is played by integrability \footnote{ For
review and extensive list of references, see
\cite{Arutyunov:2009ga,Beisert:2010jr}.}. Explicitly, the scaling
dimensions in planar $\mathcal{N}=4$ SYM can be described by the
corresponding energy levels of a string on $AdS_5\times S^5$ and
these levels can be computed using the integrability of the string,
where the spectrum of $AdS_5\times S^5$ superstring is determined by
means of the thermodynamic Bethe ansatz that is applied to a doubly
Wick rotated version of world-sheet theory
\cite{Arutyunov:2007tc,Arutyunov:2009zu,Arutyunov:2009ur}. It is
important to stress that this double Wick rotation is applied on the
light cone gauge fixed $AdS_5\times S^5$ string which is not Lorentz
invariant. As a result  the double Wick rotation leads to the
non-equivalent quantum field theory known as a mirror theory. Given
mirror theory was very useful for the description of planar
scattering amplitudes
\cite{Basso:2013vsa,Basso:2013aha,Basso:2014koa}. Then we can ask
the question whether mirror symmetry has deeper physical meaning
rather than to be considered as a technical tool. In other words we
would like to see whether the mirror theory arises by uniform gauge
fixing of a free string in some background. This was shown firstly
in \cite{Arutyunov:2014cra} for the case of the bosonic string in
$AdS_5\times S^5$ background and its integrable deformation
\cite{Delduc:2013qra,Arutyunov:2014jfa,Lunin:2014tsa,Delduc:2014kha,Hoare:2014pna},
for review, see \cite{vanTongeren:2013gva}. The analysis was
generalized to the Green-Schwarz superstring in
\cite{Arutyunov:2014jfa} where it was also suggested how the dilaton
and Ramond-Ramond fields should transform under mirror symmetry,
even if it is not completely clear whether the mirror background is
the string background as well.

It is well known that string theories contain another extended
objects in their spectrum   as for example D-branes
\cite{Polchinski:1995mt}. D1-brane has the special meaning since it
is two dimensional object that is very  similar to the fundamental
string. Further,  D1-brane couples to the dilaton and Ramond-Ramond
fields through its Dirac-Born-Infeld action and Wess-Zummino term.
For that reason the study of D1-brane in more general background
could be very useful for the analysis of the integrability of these
general backgrounds. Previously we analyzed the integrability of
D1-brane on group manifold \cite{Kluson:2014uaa} and we showed that
this theory is integrable. Now we would like to extend the analysis
of D1-branes further and try to  find uniform light cone gauge fixed
D1-brane in general background with the presence of the
Ramond-Ramond fields and dilaton, where   the uniform light cone
gauge was introduced for the string in $AdS_5\times S^5$ background
\cite{Kruczenski:2004kw,Arutyunov:2004yx}. Then we
 analyze how the gauge fixed D1-brane action
transforms under double Wick rotation and how the background fields
should transform in order to preserve the form of the gauge fixed
action. More precisely, in order to find the gauge fixed form of the
D1-brane action we firstly find the Hamiltonian formulation of
D1-brane in general background. We identify three first class
constraints where two of them correspond to the world-volume
diffeomorphism invariance and the last one to the gauge invariance
of given theory related to the presence of the gauge field
$A_\alpha$ on the world-volume of D1-brane. We fix these first class
constraints by imposing uniform light cone gauge and then we find
the Hamiltonian for the physical degrees of freedom only. Finally we
determine the Lagrangian for the physical degrees of freedom which
is the main goal of this paper. This Lagrangian could be starting
point for further  analysis of more general integrable systems, as
for example  for the analysis of giant magnon solution in
$\kappa-$deformed background, for the analysis of the world-sheet
S-matrix with the presence of Ramond-Ramond fields  and so on.
However in given article  we restrict ourselves to the question how
gauge fixed D1-brane Lagrangian behaves under double Wick rotation.
From the requirement that the double Wick rotated action should have
the same form as the original one we determine mirror background.
However due to the fact that the gauge fixed action possesses
constant electric flux that is proportional to the number of
fundamental strings we find that this flux is involved in the
transformation properties of the background fields. Of  course, this
is a trivial situation in case of  a constant dilaton and zero
Ramond-Ramond one form as in case of $AdS_5\times S^5$ background
since then we can rescale the spatial coordinate of D1-brane theory
so that the dependence on the electric flux and constant dilaton
appears as a  constant prefactor in front of the action. In other
words in case of the background with constant dilaton and zero
Ramond-Ramond fields the action behaves under the  double Wick
rotation as uniform  gauge fixed bosonic string. However there is an
important difference that makes the meaning of the double Wick
rotation in case of gauge fixed D1-brane theory unclear. It was
argued in \cite{Arutyunov:2014cra} that the double Wick rotation for
the bosonic string in $AdS_5\times S^5$ gives the mirror background
with non-trivial dilaton. On the other hand the double Wick rotation
in case of gauge fixed D1-brane action implies that
 $e^{-2\Phi}+(C^{(0)})^2$  does  not transform and hence we would
 find that non-trivial dilaton in the mirror background
 should be supported corresponding
 non-trivial Ramond-Ramond zero form. Further, when we apply
 double Wick rotation to the case of gauge fixed D1-brane in
 $\kappa$ deformed $AdS_3\times S^3$ background we find that Wick
 rotated theory cannot be equivalent to the original one due to the
 presence of non-trivial Ramond-Ramond two form
\cite{Lunin:2014tsa}.  In other words the double Wick rotation in
case of the gauge fixed D1-brane theory in general background cannot
give the action  that has formally the same form as the original
one. It is possible that the origin of this discrepancy is in the
fact that the double Wick rotation performed in case of the
fundamental string corresponds to the T-duality in $\phi$
coordinates, then performing analytic continuing $t,\phi\rightarrow
i\phi, -it$ and doing another $T-$duality in the new $\phi$
coordinate \cite{Arutyunov:2014jfa}. On the other hand it is not
completely clear how to perform T-duality transformation in case of
the uniform gauge fixed D1-brane theory. Recall that standard
treatment of T-duality transformations in case of D1-brane theory is
performed when the static gauge is imposed, for review see
\cite{Simon:2011rw}.

Let us outline our results. We derive the uniform gauge fixed
D1-brane action in general background. This action could be very
useful for the study of the integrability of this theory in
$\kappa-$deformed background with non-trivial dilaton and
Ramond-Ramond fields. It could be also very interesting to study
giant magnon solution on given D1-brane. On the other hand the
double Wick rotation in given theory is subtle and should be
clarified further.

The structure of given paper is as follows. In the next section
(\ref{second}) we perform Hamiltonian formulation of D1-brane in
general background. Then we perform uniform gauge fixing and
determine Lagrangian density for the physical degrees of freedom. In
section (\ref{third}) we perform the double Wick rotation in given
action and discuss  conditions under which the double Wick rotated
action has the same form as the original one.

\section{Uniform Gauge  Fixing for D1-brane in General Background}
\label{second} In this section we find the uniform gauge fixed form
of D1-brane in general background.  We begin with the D1-brane
action in general background that has the form
\begin{eqnarray}\label{SDbrane}
S&=&-T_{D1}\int d\tau d\sigma e^{-\Phi} \sqrt{-\det
(g_{\alpha\beta}+b_{\alpha\beta}+
(2\pi\alpha')F_{\alpha\beta})}+\nonumber \\
&+& T_{D1}\int d\tau d\sigma[
C^{(0)}(b_{\tau\sigma}+(2\pi\alpha')F_{\tau\sigma})+C_{\tau\sigma}^{(2)}]
\ ,
\nonumber \\
\end{eqnarray}
where
\begin{equation}
g_{\alpha\beta}=g_{MN}\partial_\alpha x^M\partial_\beta x^N \ ,
\quad  b_{\alpha\beta}=b_{MN}\partial_\alpha x^M\partial_\beta x^N \
, \quad  C^{(2)}_{\tau\sigma}=C_{MN}^{(2)}\partial_\tau
x^M\partial_\sigma x^N \ ,
\end{equation}
and where $x^M(\tau,\sigma)$ are embedding coordinates for D1-brane
in given background. Further,  $F_{\alpha\beta}=\partial_\alpha
A_\beta-
\partial_\beta A_\alpha$ is the field strength of the world-volume
gauge field $A_\alpha,\alpha=\tau,\sigma$. Finally $T_{D1}$ is
D1-brane tension $T_{D1}=\frac{1}{2\pi\alpha'}$.

Before we proceed to the Hamiltonian formulation of the action
(\ref{SDbrane}) it is useful to use following formula
\begin{equation}
\det (g_{\alpha\beta}+b_{\alpha\beta}+(2\pi\alpha')
F_{\alpha\beta})= \det g_{\alpha\beta}+(b_{\tau\sigma}+(2\pi\alpha')
F_{\tau\sigma})^2 \
\end{equation}
that holds in two dimensions only.  Then from the action
(\ref{SDbrane}) we find momenta conjugate to $x^M,A_\sigma$ and
$A_\tau$ respectively
\begin{eqnarray}
p_M&=&T_{D1} \frac{e^{-\Phi}}{\sqrt{-\det g
-((2\pi\alpha')F_{\tau\sigma}+
b_{\tau\sigma})^2}}\left(g_{MN}\partial_\alpha x^N g^{\alpha
\tau}\det
g+\right.\nonumber \\
&+&\left.((2\pi\alpha')F_{\tau\sigma}+b_{\tau\sigma})b_{MN}\partial_\sigma
x^N\right)+T_{D1}(C^{(0)}b_{MN}\partial_\sigma x^N+C^{(2)}_{MN}\partial_\sigma x^N) \ , \nonumber \\
\pi^\sigma&=&\frac{e^{-\Phi}T_{D1}(2\pi\alpha')((2\pi\alpha')F_{\tau\sigma}+
b_{\tau\sigma})}{\sqrt{-\det g -((2\pi\alpha')F_{\tau\sigma}+
b_{\tau\sigma})^2}}+T_{D1}(2\pi\alpha')C^{(0)}\ , \quad
\pi^\tau\approx 0 \ .  \nonumber \\
\end{eqnarray}
Using these relations we find that the bare Hamiltonian is equal to
\begin{eqnarray}
H=\int d\sigma (p_M\partial_\tau x^M+\pi^\sigma \partial_\tau
A_\sigma-\mL)= \int d\sigma \pi^\sigma\partial_\sigma A_\tau
\end{eqnarray}
while we have three primary constraints
\begin{eqnarray}
\mH_\sigma&\equiv& p_M\partial_\sigma x^M\approx 0 \ , \quad \pi^\tau\approx 0 \ ,  \nonumber \\
\mH_\tau &\equiv & \Pi_M
g^{MN}\Pi_N+\left(T_{D1}^2e^{-2\Phi}+\left(\frac{\pi^\sigma}{(2\pi\alpha')}
-T_{D1}C^{(0)}\right)^2\right)g_{MN}\partial_\sigma
x^M\partial_\sigma x^N \ ,
\nonumber \\
\end{eqnarray}
where
\begin{eqnarray}
\Pi_M&\equiv &
p_M-\frac{\pi^\sigma}{(2\pi\alpha')}b_{MN}\partial_\sigma x^N
-T_{D1}C^{(2)}_{MN}\partial_\sigma x^N \ .
\nonumber \\
\end{eqnarray}
According to the standard treatment of the constraint systems we
introduce the extended Hamiltonian with all primary constraints
included
\begin{equation}
H_E=\int d\sigma (\lambda_\tau\mH_\tau+\lambda_\sigma
\mH_\sigma-A_\tau\partial_\sigma\pi^\sigma+v_\tau \pi^\tau) \ ,
\end{equation}
where $\lambda_{\tau,\sigma}$ and $v_\tau$ are Lagrange
multipliers corresponding to the constraints $\mH_\tau,\mH_\sigma $ and
$\pi^\tau$.
Now the requirement of the preservation of the primary constraint
$\pi^\tau\approx 0$ implies the secondary constraint
\begin{equation}
\mG=\partial_\sigma \pi^\sigma\approx 0 \ .
\end{equation}
We should also check  that the constraints $\mH_\tau,\mH_\sigma$ are
preserved during the time evolution as well. Performing the same
calculations as in \cite{Kluson:2014uaa} we obtain
\begin{equation}
\pb{\mH_\tau(\sigma),\mH_\tau(\sigma')}\approx 0 \ , \pb{\mH_\tau
(\sigma),\mH_\sigma(\sigma')}\approx 0 \ ,
\pb{\mH_\sigma(\sigma),\mH_\sigma(\sigma')}\approx 0 \ .
\end{equation}
In other words $\mH_\tau\approx 0 \ , \mH_\sigma \approx 0$ and
$\mG\approx 0$ are the first class constraints. Our goal is to find
the formulation of the D1-brane action  where these first class
constraints are fixed. To be more specific let us presume that the
background has the form
\cite{Arutyunov:2013ega,Arutynov:2014ota,Arutyunov:2014cra}
\begin{eqnarray}
ds^2&=&g_{MN}dx^M
dx^N=g_{tt}dt^2+g_{\varphi\varphi}d\varphi^2+g_{\mu\nu}dx^\mu dx^\nu \ , \nonumber \\
 B&=&b_{MN}dX^M dX^N=b_{\mu\nu}dx^\mu dx^\nu \ , \nonumber \\
\end{eqnarray}
where $\mu,\nu$ denote the transverse directions.
 Following
\cite{Arutyunov:2013ega,Arutynov:2014ota} we introduce light cone
coordinates
\begin{equation}
x^-=\varphi-t \ , \quad x^+=(1-a)t+a\varphi \
\end{equation}
with inverse relations
\begin{equation}
t=x^+-ax^- \ , \quad \varphi=x^++(1-a)x^- \ .
\end{equation}
Then corresponding metric components have the form
\begin{equation}
G_{++}=g_{tt}+g_{\varphi\varphi} \ , \quad
G_{--}=g_{tt}a^2+(1-a)^2g_{\varphi\varphi} \ , \quad  G_{+-}=
-ag_{tt}+(1-a)g_{\varphi\varphi} \
\end{equation}
with inverse
\begin{eqnarray}
G^{++}&=&
\frac{g_{tt}a^2+(1-a)^2g_{\varphi\varphi}}{g_{tt}g_{\varphi\varphi}}
\ ,  \quad  G^{--}=
\frac{g_{tt}+g_{\varphi
 \varphi}}{g_{tt}g_{\varphi\varphi}} \ ,
 \nonumber \\
 G^{+-}&=&
 \frac{ag_{tt}-(1-a)g_{\varphi\varphi}}{g_{tt}g_{\varphi\varphi}} \
 .
 \nonumber \\
\end{eqnarray}
Further, using the relation between light-cone coordinates and the
original ones we obtain following components of Ramond-Ramond two form
\begin{eqnarray}
C^{(2)}_{+-}&=& -C^{(2)}_{-+}=C^{(2)}_{t\varphi} \ ,
 \nonumber \\
C^{(2)}_{+\mu}&=&-C_{\mu+}= C_{t\mu}^{(2)}+C_{\varphi\mu}^{(2)} \
,
\nonumber \\
C_{-\mu}^{(2)}&=&-C^{(2)}_{\mu -}=
-C_{t\mu}^{(2)}+(1-a)C_{\varphi\mu}^{(2)} \ . \nonumber \\
\end{eqnarray}
In the light cone coordinates the  Hamiltonian and diffeomorphism
constraints have the form
\begin{eqnarray}
\mH_\tau&=&\Pi_+G^{++}\Pi_++2\Pi_+G^{+-}\Pi_-+\Pi_-G^{--}\Pi_- +\nonumber \\
&+&
\left(T_{D1}^2e^{-2\Phi}+\left(\frac{\pi_\sigma}{2\pi\alpha'}-T_{D1}C^{(0)}\right)^2\right)
\left(G_{++}(\partial_\sigma x^+)^2+\right.\nonumber \\
&+& \left. 2G_{+-}\partial_\sigma x^+\partial_\sigma x^-+G_{--}(\partial_\sigma
x^-)^2\right)+ \mH_x \ ,
 \nonumber \\
\mH_\sigma &=&p_+\partial_\sigma x^++p_-\partial_\sigma x^-+p_\mu
\partial_\sigma x^\mu \ ,  \nonumber \\
\end{eqnarray}
where
\begin{eqnarray}
 \mH_x &\equiv & \Pi_\mu g^{\mu\nu}\Pi_\nu
+\nonumber \\
&+& T_{D1}^2e^{-2\Phi}g_{\mu\nu}\partial_\sigma x^\mu\partial_\sigma
x^\nu+
\left(\frac{1}{(2\pi\alpha')}\pi^\sigma-T_{D1}C^{(0)}\right)^2
\partial_\sigma x^\mu g_{\mu\nu}
\partial_\sigma x^\nu \ .
\nonumber \\
\end{eqnarray}
Now we are ready to impose the uniform  light cone gauge fixing
 when we introduce
two gauge fixing constraints
\begin{equation}
\mG_+\equiv x^+-\tau \approx  0 \ , \quad \mG_-\equiv  p_--J\approx
0 \ .
\end{equation}
It is now easy to see that these constraints form the second class
constraints together with $\mH_\tau,\mH_\sigma$. Further, since we
are interested in the Hamiltonian for the physical degrees of
freedom we also fix the gauge symmetry generated by the constraint
$\mG=\partial_\sigma \pi^\sigma\approx 0$. We fix this symmetry by
imposing the condition $A_\sigma=\mathrm{const}$. Then from the
equation of motion for $\pi^\sigma$ we find that
$\pi^\sigma=\pi^\sigma(\sigma)$. Finally using the strong form of
the constraint $\mG$ we  find that $\pi^\sigma=\mathrm{const}$.
 Then
the action has the form
\begin{eqnarray}
S&=&\int d\tau d\sigma (p_+\partial_\tau x^+ +p_-\partial_\tau x^-+
p_\mu\partial_\tau x^\mu -H_{fixed})=\nonumber \\
&=&\int d\tau d\sigma (p_\mu x^\mu+p_+ ) \ ,
\nonumber \\
\end{eqnarray}
where we used the fact that $H_{fixed}$ is the sum of the second class
constraints that vanish strongly. From the previous expression
 we see that we can identify the Hamiltonian for the reduced
system as
\begin{equation}
\mH_{red}=-p_+ \ .
\end{equation}
Now using $\mH_\tau=0,   \mH_\sigma=0$ and $\mG_+=0,
\mG_-=0$ we obtain
\begin{equation}
\partial_\sigma x^-=-\frac{1}{J}p_\mu\partial_\sigma x^\mu \
\end{equation}
and hence $\mH_\tau$ is equal to
\begin{eqnarray}
\mH_\tau&=&
\Pi_+G^{++}\Pi_++2\Pi_+G^{+-}\Pi_-+\Pi_-G^{--}\Pi_-+\nonumber \\
&+&
\left(T_{D1}^2e^{-2\Phi}+\left(\frac{\pi_\sigma}{2\pi\alpha'}-T_{D1}C^{(0)}\right)^2\right)
G_{--}(\partial_\sigma
x^-)^2+ \mH_x=0 \ . \nonumber \\
\end{eqnarray}
We solve this equation for $\Pi_+$ and we obtain
\begin{equation}
\Pi_+=\frac{-2G^{+-}\Pi_--\sqrt{\bK}}{2G^{++}} \ ,
 \end{equation}
where
\begin{equation}
\Pi_+=p_+-T_{D1}C^{(2)}_{+\mu}\partial_\sigma
x^\mu+\frac{T_{D1}}{J}C^{(2)} _{+-}(p_\mu\partial_\sigma x^\mu)  \ ,
\end{equation}
and where
\begin{eqnarray}
\bK&=& 4(G^{+-}\Pi_-)^2-\nonumber
\\
&-&4G^{++}\left[G^{--}\Pi_-^2+\left(T_{D1}^2e^{-2\Phi}+
\left(\frac{\pi^\sigma}{(2\pi\alpha')}-T_{D1}C^{(0)}\right)^2\right)
G_{--}\frac{1}{J^2}(p_\mu\partial_\sigma x^\mu)^2+ \mH_x\right] \ .
\nonumber \\
\end{eqnarray}
In order to simplify resulting expressions we use following
notations
\begin{eqnarray}
 \Pi_\mu&=&
p_\mu-\frac{\pi^\sigma}{(2\pi\alpha')}b_{\mu\nu}\partial_\sigma
x^\nu -T_{D1}C^{(2)}_{\mu\nu}\partial_\sigma
x^\nu+T_{D1}C^{(2)}_{\mu-}\frac{1}{J}(p_\nu\partial_\sigma
x^\nu)\equiv
 \nonumber \\
&\equiv &
 \bA_\mu^{ \ \nu}p_\nu-\bV_\mu \ , \nonumber \\
\bV_\mu &=&\frac{\pi^\sigma}{(2\pi\alpha')}b_{\mu\nu}\partial_\sigma
x^\nu+T_{D1}C^{(2)}_{\mu\nu}\partial_\sigma x^\nu
 \ , \quad  \bA_\mu^{ \ \nu}=\delta_\mu^{ \ \nu}+
\frac{T_{D1}}{J}C_{\mu -}^{(2)}\partial_\sigma x^\nu \ .  \nonumber
\\
\Pi_-&=& J-T_{D1}C_{-\mu}^{(2)}\partial_\sigma x^\mu\ . \nonumber \\
\end{eqnarray}
To proceed further note that by definition we have the relation
\begin{equation}
\partial_\sigma x^\mu p_\mu=\frac{1}{1+
\frac{T_{D1}}{J}\partial_\sigma x^\rho C_{\rho -}^{(2)}}
\partial_\sigma x^\mu \Pi_\mu \ .
\end{equation}
Using these formulas  we can write $\bK$ as
\begin{eqnarray}\label{defbK}
\bK
=4 (G^{+-}G^{+-}-G^{++}G^{--})\Pi^2_- -4G^{++}\bB^{\mu\nu}\Pi_\mu
\Pi_\nu-4G^{++}V \  , \nonumber \\
\end{eqnarray}
where
\begin{eqnarray}
A&=&\left(T_{D1}^2e^{-2\Phi}+
\left(\frac{\pi^\sigma}{(2\pi\alpha')}-T_{D1}C^{(0)}\right)^2\right)
\frac{G_{--}}{J^2 (1+ \frac{T_{D1}}{J}\partial_\sigma x^\rho
C_{\rho -}^{(2)})^2} \ , \nonumber \\
\bB^{\mu\nu}&=& A\partial_\sigma x^\mu\partial_\sigma
x^\nu+g^{\mu\nu}
\ , \nonumber \\
V&=&\left(T_{D1}^2e^{-2\Phi}+
\left(\frac{\pi^\sigma}{(2\pi\alpha')}-T_{D1}C^{(0)}\right)^2\right)g_{\sigma\sigma}
\ , \quad g_{\sigma\sigma}=g_{\mu\nu}\partial_\sigma x^\mu \partial_\sigma x^\nu \ . \nonumber \\
\end{eqnarray}
Collecting these results we derive the Hamiltonian density for the
physical degrees of freedom in the form
\begin{equation}
\mH_{red}=-p_+=-T_{D1}C_{+\mu}\partial_\sigma x^\mu+\frac{T_{D1}}{J(
1+ \frac{T_{D1}}{J}\partial_\sigma x^\rho C_{\rho
-}^{(2)})}\partial_\sigma x^\mu \Pi_\mu+\frac{G^{+-}}{G^{++}}\Pi_-
+\frac{1}{2G^{++}}\sqrt{\bK} \ .
\end{equation}
Our goal is to find the Lagrangian density for the physical degrees
of freedom. To do this we use the equation of motion for $x^\mu$
that follow from the Hamiltonian principle
\begin{eqnarray}
\partial_\tau x^\mu=\pb{x^\mu,H_{red}}=\frac{T_{D1}C_{+-}^{(2)}}{J
(1+ \frac{T_{D1}}{J}\partial_\sigma x^\rho C_{\rho
-}^{(2)})}\partial_\sigma x^\nu \bA_\nu^{ \ \mu}-\frac{1}{\sqrt{\bK}}
2 \Pi_\rho \bB^{\rho\nu}\bA_\nu^{ \ \mu} \ . \nonumber \\
\end{eqnarray}
We can invert this equation and we obtain
\begin{equation}\label{Pimu}
\Pi_\mu=-\frac{\sqrt{\bK}}{2} \bX^\nu (\bA^{-1})_\nu^{ \ \rho}
\bB^{-1}_{\rho\mu} \ ,
\end{equation}
where
\begin{eqnarray}\label{Xinver}
\bX^\mu&=&\partial_\tau x^\mu- \frac{T_{D1}C_{+-}^{(2)}}{J (1+
\frac{T_{D1}}{J}\partial_\sigma x^\rho C_{\rho
-}^{(2)})}\partial_\sigma x^\nu \bA_\nu^{ \ \mu} \ , \nonumber \\
\bB_{\mu\nu}^{-1}&=& g_{\mu\nu}-A\frac{g_{\mu\sigma}\partial_\sigma
x^\sigma \partial_\sigma x^\omega g_{\omega\nu}} {1+A
g_{\sigma\sigma}} \ , \nonumber \\
(\bA^{-1})_\mu^{ \ \nu}&=&\delta_\mu^{ \ \nu}-
\frac{T_{D1}}{J}\frac{ C_{\mu -}\partial_\sigma
x^\nu}{1+\frac{T_{D1}}{J} \partial_\sigma x^\mu C_{\mu -}}  \ .
\nonumber \\
\end{eqnarray}
Using (\ref{Xinver}) we obtain the Lagrangian density for the
physical degrees of freedom
\begin{eqnarray}
\mL_{red}=p_\mu\partial_\tau x^\mu -\mH_{red}
=\left(\bX^\nu+ \frac{T_{D1}C_{+-}^{(2)}}{J (1+
\frac{T_{D1}}{J}\partial_\sigma x^\rho C_{\rho
-}^{(2)})}\partial_\sigma x^\sigma \bA_\sigma^{ \  \nu
}\right)(\bA^{-1})_\nu^{ \ \mu}\bV_\mu+\nonumber
\\
+T_{D1}C_{+\mu}\partial_\sigma x^\mu-\frac{G^{+-}}{G^{++}}\Pi_- +
\frac{2}{G^{++}\sqrt{\bK}} (G^{++}V-(G^{+-}G^{+-}-G^{++}G^{--})) \ .
\nonumber \\
\end{eqnarray}
However given Lagrangian still depends on conjugate momenta $p_\mu$
through the function $\bK$. In order to express $\bK$ as the
function of time derivatives of $x^\mu$ we insert (\ref{Pimu}) into
the definition of $\bK$ (\ref{defbK}) and we obtain
\begin{equation}
\bK=\frac{4}{1+G^{++}\bX^\rho(\bA^{-1})_\rho^{ \ \mu} \bB_{\mu\nu}
\bX^\omega (\bA^{-1})_\omega^{ \
\nu}}((G^{+-}G^{+-}-G^{++}G^{--})\Pi_-^2-G^{++}V) \
\end{equation}
so that the Lagrangian density $\mL_{red}$ has the form
\begin{eqnarray}
\mL_{red}&=& \left(\bX^\mu+ \frac{T_{D1}C_{+-}^{(2)}}{J (1+
\frac{T_{D1}}{J}\partial_\sigma x^\rho C_{\rho
-}^{(2)})}\partial_\sigma x^\nu \bA_\nu^{ \
\mu}\right)(\bA^{-1})_\nu^{ \ \mu}\bV_\mu+\nonumber
\\
&+&T_{D1}C_{+\mu}\partial_\sigma x^\mu-\frac{G^{+-}}{G^{++}}\Pi_-
-\nonumber \\
&-& \frac{1}{G^{++}}\sqrt{-( 1+G^{++}\bX^\rho(\bA^{-1})_\rho^{ \ \mu}
\bB_{\mu\nu} \bX^\omega (\bA^{-1})_\omega^{ \ \nu})
(G^{++}V-(G^{+-}G^{+-}-G^{++}G^{--}))} \ .
\nonumber \\
\end{eqnarray}
 We  simplify given expression  using
\begin{eqnarray}
& &\left(\bX^\mu+ \frac{T_{D1}C_{+-}^{(2)}}{J (1+
\frac{T_{D1}}{J}\partial_\sigma x^\rho C_{\rho
-}^{(2)})}\partial_\sigma x^\nu \bA_\nu^{ \
\mu}\right)(\bA^{-1})_\nu^{ \ \mu}\bV_\mu=
\nonumber \\
& &=\frac{\pi^\sigma}{2\pi\alpha'} \partial_\tau x^\mu b_{\mu\nu}
\partial_\sigma x^\nu+T_{D1}\partial_\tau x^\mu C_{\mu\nu}^{(2)}
\partial_\sigma x^\nu
\   \nonumber \\
\end{eqnarray}
together with
\begin{eqnarray}
& &(\bX \bA^{-1})^\mu \bB_{\mu\nu}^{-1}(\bX\bA^{-1})^\nu=
\frac{1}{1+Ag_{\sigma\sigma}} \times
\nonumber \\
&\times & \left[1+A g_{\sigma\sigma}
+G^{++}g_{\tau\tau}+AG^{++}(g_{\tau\tau}g_{\sigma\sigma}-g_{\tau\sigma}
g_{\tau\sigma})- \right.\nonumber
\\
&-&2g_{\tau\sigma}G^{++} \frac{T_{D1}}{J
(1+\frac{T_{D1}}{J}\partial_\sigma x^\mu C_{\mu -}^{(2)})}
(\partial_\tau x^\mu C_{\mu -}^{(2)}+C_{+-}) +\nonumber
\\
&+& \left. g_{\sigma\sigma}G^{++} \left(\frac{T_{D1}}{J
(1+\frac{T_{D1}}{J}\partial_\sigma x^\mu C_{\mu -}^{(2)})}
(\partial_\tau x^\mu C_{\mu -}^{(2)}+C_{+-})\right)^2\right]
\nonumber \\
\end{eqnarray}
Finally using the fact that we can write
\begin{eqnarray}
G^{++}V+(G^{++}G^{--}-G^{+-}G^{+-})\Pi_-^2=\nonumber \\
=(G^{++}G^{--}-G^{+-}G^{+-})\Pi_-^2 (1+Ag_{\sigma\sigma}) \nonumber
\\
\end{eqnarray}
we obtain the light-cone gauge fixed Lagrangian density for D1-brane
in the form
\begin{eqnarray}
\mL_{red}&=&\frac{\pi^\sigma}{2\pi\alpha'} \partial_\tau x^\mu
b_{\mu\nu}
\partial_\sigma x^\nu+T_{D1}\partial_\tau x^\mu C_{\mu\nu}^{(2)}
\partial_\sigma x^\nu
 +T_{D1}C^{(2)}_{+\mu}\partial_\sigma x^\mu-\nonumber \\
 &-&\frac{G^{+-}}{G^{++}}(J+T_{D1}C_{\mu -}^{(2)}\partial_\sigma x^\mu)
-\sqrt{-\frac{1}{G^{++}G_{--}}(J+T_{D1}C_{\mu -}^{(2)}\partial_\sigma x^\mu)^2
\mK} \ , \nonumber \\
\end{eqnarray}
where
\begin{eqnarray}
\mK&=& 1+ \left(e^{-2\Phi}+ \left(\pi^\sigma-C^{(0)}\right)^2\right)
\frac{T_{D1}^2}{J^2}\frac{G_{--}}{ (1+
\frac{T_{D1}}{J}\partial_\sigma x^\rho C_{\rho -}^{(2)})^2}
  g_{\sigma\sigma}+\nonumber \\
&+&G^{++}g_{\tau\tau}+ \left(e^{-2\Phi}+
\left(\pi^\sigma-C^{(0)}\right)^2\right) \frac{T_{D1}^2}{J^2}
\frac{G^{++}G_{--}}{ (1+ \frac{T_{D1}}{J}\partial_\sigma x^\rho
C_{\rho -}^{(2)})^2} (g_{\tau\tau}g_{\sigma\sigma}-g_{\tau\sigma}
g_{\tau\sigma})- \nonumber
\\
&-& 2\frac{T_{D1}}{J}g_{\tau\sigma}G^{++} \frac{1}{
1+\frac{T_{D1}}{J}\partial_\sigma x^\mu C_{\mu -}^{(2)}}
(\partial_\tau x^\mu C_{\mu -}^{(2)}+C^{(2)}_{+-}) +\nonumber
\\
&+&\frac{T_{D1}^2}{J^2}g_{\sigma\sigma}G^{++} \left(\frac{1}{
1+\frac{T_{D1}}{J}\partial_\sigma x^\mu C_{\mu -}^{(2)}}
(\partial_\tau x^\mu C_{\mu -}^{(2)}+C^{(2)}_{+-})\right)^2 \ , \nonumber \\
\end{eqnarray}
and where we also used the fact that  $T_{D1}=\frac{1}{2\pi\alpha'}$. If we now
perform rescaling the spatial coordinate as
\begin{equation}
\frac{J}{T_{D1}}\sigma =\sigma' \ ,
\end{equation}
we obtain the action in the form
\begin{equation}
S=\int d\sigma' d\tau \mL_{red} \ ,
\end{equation}
where
\begin{eqnarray}\label{mLfinal}
\mL_{red} &=&\frac{\pi^\sigma}{2\pi\alpha'} \partial_\tau x^\mu
b_{\mu\nu}
\partial_{\sigma'} x^\nu+T_{D1}\partial_\tau x^\mu C_{\mu\nu}^{(2)}
\partial_{\sigma '} x^\nu
 +T_{D1}C^{(2)}_{+\mu}\partial_{\sigma'} x^\mu-\nonumber \\
&-& T_{D1}\frac{G^{+-}}{G^{++}}(1+C_{\mu -}^{(2)}\partial_{\sigma'}
x^\mu) -T_{D1}\sqrt{-\frac{1}{G^{++}G_{--}}(1+C_{\mu
-}^{(2)}\partial_{\sigma'} x^\mu)^2
\mK} \ , \nonumber \\
\mK&=& 1+ \left(e^{-2\Phi}+ \left(\pi^\sigma-C^{(0)}\right)^2\right)
\frac{G_{--}}{ (1+ \partial_{\sigma'} x^\rho C_{\rho -}^{(2)})^2}
  g_{\sigma\sigma}+\nonumber \\
&+& G^{++}g_{\tau\tau}+ \left(e^{-2\Phi}+
\left(\pi^\sigma-C^{(0)}\right)^2\right) \frac{G^{++}G_{--}}{ (1+
\partial_{\sigma'} x^\rho C_{\rho -}^{(2)})^2}
(g_{\tau\tau}g_{\sigma\sigma}-g_{\tau\sigma} g_{\tau\sigma})-
\nonumber
\\
&-&2g_{\tau\sigma}G^{++} \frac{1}{ 1+\partial_{\sigma'} x^\mu C_{\mu
-}^{(2)}} (\partial_\tau x^\mu C_{\mu -}^{(2)}+C^{(2)}_{+-})
+\nonumber
\\
&+& g_{\sigma\sigma}G^{++} \left(\frac{1}{ 1+\partial_{\sigma'}
x^\mu C_{\mu -}^{(2)}} (\partial_\tau x^\mu C_{\mu
-}^{(2)}+C^{(2)}_{+-})\right)^2 \ .
\nonumber \\
\end{eqnarray}
This is the final form of gauge fixed D1-brane in general
background. It is easy to see that when Ramond-Ramond fields vanish,
$\Phi=\mathrm{const}$  and when $\pi^\sigma=0$ the action has
formally the same form as the gauge-fixed form of the fundamental
string action in arbitrary background $g_{MN},b_{MN}$ with vanishing
light-cone components of $B$ that was derived in
\cite{Arutynov:2014ota}. Of course, strictly speaking these two
actions coincide in the limit of infinite long string since we
started with D1-brane that is infinite extended.

\section{Double Wick Rotation}\label{third}
In this section we perform double Wick rotation in case of the gauge
fixed Lagrangian density given in (\ref{mLfinal}) following very
nice analysis performed in \cite{Arutyunov:2014cra} where also very
interesting conjecture related to the double Wick rotation was
suggested. It was argued there that  due to the fact that the light
cone gauge fixed $AdS_5\times S^5$ string is not Lorentz invariant
we obtain that the double Wick rotation gives an inequivalent
theory, so called mirror theory. Since this mirror symmetry is
important in the study of the integrability of the string in
$AdS_5\times S^5$ it was suggested in \cite{Arutyunov:2014cra} that
the double Wick rotation could be fundamental in the definition of
the mirror  theory. Explicitly, it was shown in
\cite{Arutyunov:2014cra} that the application of the double Wick
rotation to the case of the light cone gauge fixed string on some
background gives the string theory formulated on another background.

Due to the  similarity between fundamental string action and
D1-brane action we would like to extend given idea  to the case of
the light-cone gauge fixed D1-brane action introduced above. Clearly
given action is more complicated due to the explicit coupling of
D1-brane to the dilaton and Ramond-Ramond fields which is invisible
in the classical bosonic string. For that reason it is instructive
to see how the action (\ref{mLfinal}) changes under  double Wick
rotation that is defined as
\begin{eqnarray}
\tau=i\tsigma \ , \quad \sigma'=-i\ttau
\end{eqnarray}
so that
\begin{eqnarray}
\frac{\partial}{\partial \tau}=-i\frac{\partial}{\partial \tsigma}
\quad  , \frac{\partial}{\partial \sigma'}=i\frac{\partial}{\partial
\ttau} \
. \nonumber \\
\end{eqnarray}
Now under these transformations $\mK$ takes the form
\begin{eqnarray}
\mK&=&1- \left(e^{-2\Phi}+ \left(\pi^\sigma-C^{(0)}\right)^2\right)
\frac{G_{--}}{ (1+ i\partial_{\tsigma} x^\rho C_{\rho -}^{(2)})^2}
  g_{\ttau\ttau}-\nonumber \\
&-&G^{++}g_{\tsigma\tsigma}+ \left(e^{-2\Phi}+
\left(\pi^\sigma-C^{(0)}\right)^2\right) \frac{G^{++}G_{--}}{ (1+
i\partial_{\ttau} x^\rho C_{\rho -}^{(2)})^2}
(g_{\ttau\ttau}g_{\tsigma\tsigma}-g_{\ttau\tsigma}
g_{\ttau\tsigma})- \nonumber
\\
&-&2g_{\ttau\tsigma}G^{++} \frac{1}{ (1+i\partial_{\ttau} x^\mu
C_{\mu -}^{(2)})} (-i\partial_{\ttau} x^\mu C_{\mu -}^{(2)}+C_{+-})
+\nonumber\\
&-&g_{\ttau\ttau}G^{++} \left(\frac{1}{ (1+i\partial_{\ttau} x^\mu
C_{\mu -}^{(2)})} (-\partial_{\tsigma} x^\mu C_{\mu
-}^{(2)}+C_{+-})\right)^2
\end{eqnarray}
We see that in order to have the same form of the Lagrangian as the original one
we should  demand that
\begin{equation}\label{Cnonequval}
C^{(2)}_{\rho -}=0 \
\end{equation}
and we return to this important restriction letter.
 Let us denote the
background fields in the mirror theory with tilde. Then in order to
have mirror theory in the same form as the original one we obtain
following relations between mirror background fields and original
ones
\begin{eqnarray}
& & e^{-2\Phi}+(\pi^\sigma-C^{(0)})^2=
e^{-2\tilde{\Phi}}+(\pi^\sigma-\tilde{C}^{(0)})^2 \ , \nonumber \\
\tilde{G}^{++}&=&-(e^{-2\Phi}+(\pi^\sigma-C^{(0)})^2)G_{--}-G^{++}C^2_{+-}
\ ,
\nonumber \\
\tilde{G}_{--}&=&
-\frac{G^{++}G_{--}}{(e^{-2\Phi}+(\pi^\sigma-C^{(0)})^2)G_{--}+G^{++}C^2_{+-}}
\ ,
\nonumber \\
\tilde{C}_{+-}&=&
-\frac{G^{++}C_{+-}}{(e^{-2\Phi}+(\pi^\sigma-C^{(0)})^2)G_{--}+G^{++}C_{+-}^2}
\nonumber \\
\tilde{G}^{+-}&=& -\frac{G^{+-}}{G^{++}}
(e^{-2\Phi}+(\pi^\sigma-C^{(0)})^2)G_{--}-G^{+-}C^2_{+-} \ ,
\nonumber
\\
b_{\mu\nu}&=&-\tilde{b}_{\mu\nu} \ , \quad C_{\mu\nu}^{(2)}=-
\tilde{C}_{\mu\nu}^{(2)} \ . \nonumber \\
\end{eqnarray}
Let us now comments given results. The most important one is on the
first line that shows that we are not able to determine the
transformation properties of $\Phi$ and $C^{(0)}$ separately while
their combination that includes the world-volume electric flux is
constant. This is the first issue since the double Wick rotation
seems to mix background fields with the world-volume ones. Secondly,
even in the case of the simplest background $AdS_5\times S^5$  this
result seems to predict different transformations of dilaton from
the result  determined in \cite{Arutyunov:2014cra}. Of course, there
is still a possibility that non-trivial dilaton dependence in the
mirror theory is compensated by non-trivial behavior of $C^{(0)}$ so
that their combination is constant but it seems that such
configurations do not solve the supergravity equations of motion.
 Most importantly we see from
(\ref{Cnonequval}) that the mirror theory has the same form as the
original one when we have to impose the restriction on the
background  that $C^{(2)}_{-\mu}=0$. However it is easy to see that
this condition certainly cannot be obeyed in case of $\kappa-$
deformed $AdS_3\times S^3$ background that was found in
\cite{Lunin:2014tsa}. For these reasons we mean that the double Wick
rotation in case of gauge fixed D1-brane action generally gives
theory with different Lagrangian density so that it does not make
sense to speak about Wick rotated D1-brane theory as  mirror theory.
It is possible that the reason why D1-brane theory behaves
differently from the fundamental string action is in the fact that
the double Wick rotation performed in case of the fundamental string
corresponds to the T-duality in $\phi$ coordinate, then performing
analytic continuing $t,\phi\rightarrow i\phi, -it$ and doing another
$T-$duality in the new $\phi$ coordinate \cite{Arutyunov:2014jfa}.
On the other hand it is not known how to perform  such a T-duality
transformations in case of the uniform gauge fixed D1-brane action
however it is possible that such an operation is not equivalent to
the double Wick rotation performed above. We hope to return to this
problem in future.

 \noindent {\bf
Acknowledgement:}

 This work   was
supported by the Grant agency of the Czech republic under the grant
P201/12/G028. \vskip 5mm


\begin{thebibliography}{20}




\bibitem{Arutyunov:2009ga}
  G.~Arutyunov and S.~Frolov,
\emph{``Foundations of the $AdS_5 x S^5$ Superstring. Part I,''}
  J.\ Phys.\ A {\bf 42} (2009) 254003
  [arXiv:0901.4937 [hep-th]].


\bibitem{Beisert:2010jr}
  N.~Beisert, C.~Ahn, L.~F.~Alday, Z.~Bajnok, J.~M.~Drummond, L.~Freyhult, N.~Gromov and R.~A.~Janik {\it et al.},
\emph{``Review of AdS/CFT Integrability: An Overview,''}
  Lett.\ Math.\ Phys.\  {\bf 99} (2012) 3
  [arXiv:1012.3982 [hep-th]].

\bibitem{Kluson:2014uaa}
  J.~Kluson,
\emph{``Integrability of
 D1-brane on Group Manifold,''}
  JHEP {\bf 1409} (2014) 159
  [arXiv:1407.7665 [hep-th]].





\bibitem{Arutyunov:2013ega}
  G.~Arutyunov, R.~Borsato and S.~Frolov,
\emph{``S-matrix for strings on $\eta$-deformed $AdS_{5} \times
S^5$,''}
  JHEP {\bf 1404} (2014) 002
  [arXiv:1312.3542 [hep-th]].

\bibitem{Arutynov:2014ota}
  G.~Arutyunov, M.~de Leeuw and S.~J.~van Tongeren,
\emph{``The exact spectrum and mirror duality of the ($AdS_{5}
\times$ S$^5$)$_{\eta}$ superstring,''}
  Theor.\ Math.\ Phys.\  {\bf 182} (2015) 1,  23
   [Teor.\ Mat.\ Fiz.\  {\bf 182} (2014) 1,  28]
  [arXiv:1403.6104 [hep-th]].

\bibitem{Arutyunov:2014cra}
  G.~Arutyunov and S.~J.~van Tongeren,
\emph{``$\mathrm{AdS}_5 \times \mathrm{S}^5$ mirror model as a
string sigma model,''}
  Phys.\ Rev.\ Lett.\  {\bf 113}, no. 26, 261605 (2014)
  [arXiv:1406.2304 [hep-th]].


\bibitem{Arutyunov:2007tc}
  G.~Arutyunov and S.~Frolov,
\emph{``On String S-matrix, Bound States and TBA,''}
  JHEP {\bf 0712} (2007) 024
  [arXiv:0710.1568 [hep-th]].


\bibitem{Polchinski:1995mt}
  J.~Polchinski,
\emph{``Dirichlet Branes and Ramond-Ramond charges,''}
  Phys.\ Rev.\ Lett.\  {\bf 75} (1995) 4724
  [hep-th/9510017].


\bibitem{Arutyunov:2009zu}
  G.~Arutyunov and S.~Frolov,
\emph{``String hypothesis for
 the AdS(5) x S**5 mirror,''}
  JHEP {\bf 0903} (2009) 152
  [arXiv:0901.1417 [hep-th]].

\bibitem{Arutyunov:2009ur}
  G.~Arutyunov and S.~Frolov,
\emph{``Thermodynamic Bethe Ansatz for the AdS(5) x S(5) Mirror
Model,''}
  JHEP {\bf 0905} (2009) 068
  [arXiv:0903.0141 [hep-th]].



\bibitem{Basso:2013vsa}
  B.~Basso, A.~Sever and P.~Vieira,
\emph{``Spacetime and Flux Tube S-Matrices at Finite Coupling for
N=4 Supersymmetric Yang-Mills Theory,''}
  Phys.\ Rev.\ Lett.\  {\bf 111} (2013) 9,  091602
  [arXiv:1303.1396 [hep-th]].

\bibitem{Basso:2013aha}
  B.~Basso, A.~Sever and P.~Vieira,
\emph{``Space-time S-matrix and Flux tube S-matrix II. Extracting
and Matching Data,''}
  JHEP {\bf 1401} (2014) 008
  [arXiv:1306.2058 [hep-th]].

\bibitem{Basso:2014koa}
  B.~Basso, A.~Sever and P.~Vieira,
\emph{``Space-time S-matrix and Flux-tube S-matrix III. The
two-particle contributions,''}
  JHEP {\bf 1408} (2014) 085
  [arXiv:1402.3307 [hep-th]].



\bibitem{Delduc:2013qra}
  F.~Delduc, M.~Magro and B.~Vicedo,
\emph{``An integrable deformation of the $AdS_5 \times S^5$
superstring action,''}
  Phys.\ Rev.\ Lett.\  {\bf 112} (2014) 5,  051601
  [arXiv:1309.5850 [hep-th]].

\bibitem{Arutyunov:2014jfa}
  G.~Arutyunov and S.~J.~van Tongeren,
\emph{``Double Wick rotating Green-Schwarz strings,''}
  arXiv:1412.5137 [hep-th].


\bibitem{Lunin:2014tsa}
  O.~Lunin, R.~Roiban and A.~A.~Tseytlin,
\emph{``Supergravity backgrounds for deformations of AdS$_{n} \times
S^n$ supercoset string models,''}
  Nucl.\ Phys.\ B {\bf 891} (2015) 106
  [arXiv:1411.1066 [hep-th]].


\bibitem{Delduc:2014kha}
  F.~Delduc, M.~Magro and B.~Vicedo,
\emph{``Derivation of the action and symmetries of the $q$-deformed
$AdS_{5} \times S^{5}$ superstring,''}
  JHEP {\bf 1410} (2014) 132
  [arXiv:1406.6286 [hep-th]].

\bibitem{Hoare:2014pna}
  B.~Hoare, R.~Roiban and A.~A.~Tseytlin,
\emph{``On deformations of $AdS_n$ x $S^n$ supercosets,''}
  JHEP {\bf 1406} (2014) 002
  [arXiv:1403.5517 [hep-th]].

\bibitem{vanTongeren:2013gva}
  S.~J.~van Tongeren,
\emph{``Integrability of the ${\rm Ad}{{{\rm S}}_{5}}\times {{{\rm
S}}^{5}}$ superstring and its deformations,''}
  J.\ Phys.\ A {\bf 47} (2014) 43,  433001
  [arXiv:1310.4854 [hep-th]].



\bibitem{Simon:2011rw}
  J.~Simon,
\emph{``Brane Effective Actions, Kappa-Symmetry and Applications,''}
  Living Rev.\ Rel.\  {\bf 15} (2012) 3
  [arXiv:1110.2422 [hep-th]].



\bibitem{Kruczenski:2004kw}
  M.~Kruczenski, A.~V.~Ryzhov and A.~A.~Tseytlin,
\emph{``Large spin limit of AdS(5) x S**5 string theory and
low-energy expansion of ferromagnetic spin chains,''}
  Nucl.\ Phys.\ B {\bf 692} (2004) 3
  [hep-th/0403120].

\bibitem{Arutyunov:2004yx}
  G.~Arutyunov and S.~Frolov,
\emph{``Integrable Hamiltonian for classical strings on AdS(5) x
S**5,''}
  JHEP {\bf 0502} (2005) 059
  [hep-th/0411089].



\end{thebibliography}
\end{document}